# Reconstruction of missing low-angle scattering in two-dimensional diffraction signal

Yanwei Xiong*, Martin Centurion†

Department of Physics and Astronomy, University of Nebraska-Lincoln, Lincoln, Nebraska, 68588, USA

**ABSTRACT**. Anisotropic two-dimensional diffraction signals encode additional structural information, including atom-pair angular distributions, beyond conventional isotropic scattering. However, experimental constraints such as beam stops result in missing low-angle scattering data, which limits accurate real-space reconstruction. We develop an iterative algorithm to recover the missing low-angle signal in two-dimensional diffraction patterns. The method transforms between momentum-transfer and real-space domains using coupled Fourier and Abel transforms, while enforcing real-space support constraints to suppress reconstruction artifacts. Importantly, the algorithm requires only minimal *a priori* knowledge of the molecular structure, namely the approximate shortest and longest internuclear distances. We demonstrate accurate reconstruction of the missing signal using both simulated data and experimental diffraction patterns from laser-aligned trifluoroiodomethane ($CF_3I$) molecules, enabling improved real-space structural retrieval from incomplete diffraction data. Our results remove a fundamental experimental limitation in ultrafast diffraction and establish a general route toward complete structural retrieval from incomplete scattering data.

## I. INTRODUCTION

Gas-phase ultrafast electron diffraction (GUED) and ultrafast X-ray diffraction (UXRD) are powerful techniques for resolving molecular structure and nuclear dynamics with femtosecond temporal and sub-angstrom spatial resolution [1, 2]. In these pump–probe experiments, a laser pulse initiates chemical reactions, and a delayed electron or X-ray pulse probes the evolving structure and records a sequence of two-dimensional (2-D) diffraction patterns, providing access to time-resolved molecular structure through real-space reconstruction or comparison with theoretical models. The small interaction cross section and low molecular density in gas-phase samples result in most incident electrons or photons being transmitted without scattering. A beam stop is therefore used to block the intense transmitted beam and protect the detector, enabling measurement of the weaker scattered signal. This configuration systematically removes diffraction signals at low scattering angles. At the same time, the maximum accessible scattering angle is limited by detector size or the signal level of the rapid decay of scattering intensity. As a result, the measurable momentum transfer range is bounded between a minimum value $s_{min}$ and a maximum value $s_{max}$. The finite $s_{max}$ can be treated using standard approaches such as damping functions. In contrast, the missing low-angle signal is critical, as it encodes long-range structural information and is essential for accurate recovery of the real-space pair distribution function (PDF).

Analysis of GUED and UXRD signals has largely focused on one-dimensional (1-D) representations corresponding to isotropic scattering in both momentum-transfer and real-space domains. Various approaches have been developed to address the missing low-angle isotropic signal, including interpolation-based methods [3, 4], model-based reconstruction using simulations [5-9], momentum-space analysis [10-13], and structure retrieval with genetic algorithm techniques [14-17]. However, these approaches often rely on strong model assumptions, require extensive prior knowledge, or lack robustness in complex systems. Recently, we introduced an iterative reconstruction algorithm for isotropic scattering [18], inspired by phase retrieval methods [19-24]. The algorithm iteratively transforms between momentum-transfer and real-space domains while enforcing support constraints to recover the missing low-angle signal. Compared to existing approaches, it is simple to implement, requires minimal *a priori* knowledge of the molecular structure, and can be applied to systems with multiple reaction pathways. The method can also be used to separate inelastic and elastic scattering contributions in GUED signals.

*Contact author: yxiong3@unl.edu
†Contact author: martin.centurion@unl.edu

Although diffraction analysis has largely focused on isotropic scattering, anisotropic 2-D diffraction patterns are routinely observed in GUED and UXRD experiments due to the use of linearly polarized excitation pulses as a pump [25]. These anisotropic signals contain additional structural information beyond isotropic scattering. For example, 2-D GUED signals enable extraction of detailed structural and dynamical information, including bond distances and angles [26-28], atom-pair angular distributions [29-31], and transient molecular structures [32-34] and reaction dynamics [35], in laser-induced processes.

Real-space reconstruction of isotropic scattering relies on a 1-D Fourier–sine transform, whereas anisotropic 2-D diffraction requires combined Fourier and Abel inversions [30, 33]. As a result, reconstruction methods developed for isotropic signals cannot be directly extended to anisotropic 2-D diffraction due to increased complexity of the inversion procedure. In this work, we extend the iterative reconstruction approach to anisotropic 2-D diffraction signal by combining the iterative reconstruction algorithm with the Fourier–Abel inversion framework. We validate the method using both simulated data and experimental electron diffraction measurements of $CF_3I$, demonstrating accurate recovery of the missing low-angle scattering signal.

## II. THEORY

In this section, we briefly review electron scattering theory for gas-phase molecules and then present the iterative algorithm for recovering the inaccessible low-angle signal in 2-D diffraction data.

### A. Electron scattering theory

Electron scattering from isolated molecules has been described extensively in previous work [7, 36-39] Elastic scattering from neutral molecules is commonly modeled using the independent atom model, in which each atom is treated as a spherically symmetric scatterer and bonding effects are neglected. In electron scattering from isolated molecules, the scattered electron waves from different atoms interfere, giving rise to the total elastic scattering intensity. The elastic scattering intensity from a molecule consisting of $N$ atoms is given by

$$I(\boldsymbol{s}) = \sum_{j=1}^{N} \sum_{k=1}^{N} f_j^*(s) f_k(s) e^{i\boldsymbol{s}\cdot\boldsymbol{r}_{jk}}, \tag{1}$$

where $f_j(s)$ denotes the atomic scattering amplitude of the atom $j$ , and $\boldsymbol{r}_{jk}$ is a vector pointing from the atom $k$ to the atom $j$ , with magnitude equal to the internuclear distance, and $\boldsymbol{s}$ is the momentum transfer with magnitude equal to $s = \frac{4\pi}{\lambda}\sin\left(\frac{\theta}{2}\right)$, in which $\lambda$ is the wavelength and $\theta$ is the scattered angle of electrons.

The total elastic diffraction intensity from a gas-phase sample is an incoherent sum of the scattering contributions from all molecules in the ensemble. We consider 2-D diffraction patterns from molecules excited by a linearly polarized laser pulse. The total scattering intensity $I_{total}(\boldsymbol{s})$ can be decomposed into an atomic term $I_A(s)$, which contains no structural information, and a molecular term $I_M(\boldsymbol{s})$, which encodes the molecular geometry.

$$I_{total}(\boldsymbol{s}) = I_A(s) + I_M(\boldsymbol{s}), \tag{2}$$

$$I_A(s) = \sum_{j=1}^{N} \left| f_j(s) \right|^2, \tag{3}$$

$$I_M(\boldsymbol{s}) = \sum_{j=1}^{N} \sum_{k=1,k\neq j}^{N} f_j^*(s) f_k(s) \iint e^{-i\boldsymbol{s}\cdot\boldsymbol{r}_{jk}(\alpha,\beta)} g_{jk}(\alpha) \sin\alpha \, d\alpha d\beta \ , \tag{4}$$

where $\alpha, \beta$ are polar and azimuthal angles that describe the orientation of the atom pair $jk$ in the laboratory frame, and $g_{jk}(\alpha)$ denotes its angular distribution from a linearly polarized laser excitation. The general relation between the atom-pair angular distribution and the molecular orientation probability density, expressed in terms of the Euler angles, is given in Refs.[31].

When the molecules are randomly oriented, the atom-pair angular distribution becomes $g_{jk}(\alpha) = 1/4\pi$, and Eq. (4) reduces to the Debye scattering equation [40, 41], formulated as $I_M^{randon}(s) = \sum_{j=1}^{N} \sum_{k \neq j}^{N} f_j^*(s) f_k(s) \sin(s\, r_{\text{jk}})/(s\, r_{\text{jk}})$, which corresponds to the isotropic scattering signal. In this case, the diffraction pattern consists of concentric rings whose intensity decreases rapidly with increasing s. However, in the general case, the integral in Eq. (4) does not admit a closed-form analytical expression.

The atomic scattering amplitudes $f_j(s)$ decrease approximately as $s^{-2}$ with increasing momentum transfer. To remove this rapid decay and enhance oscillatory features at high $s$, we define the rescaled molecular scattering intensity as $I_M(\boldsymbol{s})/I_A(s)$. The atom-pair distances and angular distributions in real space can be retrieved from the modified pair distribution function (MPDF), obtained by applying a 2-D Fourier inversion ($\mathcal{F}_{2D}^{-1}$) followed by an Abel inversion ($\mathcal{A}^{-1}$) to $I_M(\boldsymbol{s})/I_A(s)$ [31],

$$\text{MPDF} = \mathcal{A}^{-1} \mathcal{F}_{2D}^{-1} \left[ \frac{I_M(\boldsymbol{s})}{I_{\text{A}}(s)} \right] = \sum_{j=1}^{N} \sum_{k=1, j \neq k}^{N} g_{jk}(\alpha)\, \delta(r - r_{jk}) \otimes \frac{F_j(r) \star F_k(r)}{{r_{jk}}^2}, \tag{5}$$

where $\boldsymbol{s} = (s_x, s_y)$, $\otimes$ signifies convolution, $\star$ stands for correlation, $\delta(r - r_{jk})$ is the ideal PDF, and $F_j(r)$ is the Fourier transform of the normalized atomic scattering amplitude $f_j(s)/\sqrt{I_{\text{A}}(s)}$. The MPDF can be understood as an angularly dependent pair distribution function.

In time-resolved experiments, the transient signal is isolated using the diffraction-difference method, defined as the difference in total scattering intensity,

$$\Delta I_M(\boldsymbol{s}, t) = I_{total}(\boldsymbol{s}, t) - I_{total}(\boldsymbol{s}, t_0), \tag{6}$$

where $t$ is the time delay between the probe electron pulse and the laser excitation, and $t_0$ denotes a reference time before excitation. The corresponding rescaled molecular signal is given by $\Delta I_M(\boldsymbol{s})/I_A(s)$. The difference modified pair distribution function (ΔMPDF) is then obtained by substituting $I_M(\boldsymbol{s})/I_A(s)$ in Eq. (5) with $\Delta I_M(\boldsymbol{s})/I_A(s)$.

### B. Retrieval method

Here we describe an iterative algorithm for restoring missing low-$s$ signals in 2-D diffraction patterns. First, we define the transform pairs linking the momentum-transfer and real-space domains. Second, we model the measured diffraction signal over a finite range from $s_{min}$ to $s_{max}$ and analyze the artifacts introduced in real space by the missing data. Third, we define the real-space constraints to iteratively suppress the artifacts.

The rescaled molecular scattering intensity truncated at a maximum momentum transfer $s_{max}$ is defined as

$$\mathcal{M}(\boldsymbol{s}) = \begin{cases} \sum_{j=1}^{N} \sum_{k \neq j}^{N} c_{jk}(s) \iint e^{-i\boldsymbol{s} \cdot \boldsymbol{r}_{jk}(\alpha,\beta)} g_{jk}(\alpha) \sin\alpha \, d\alpha d\beta & \text{for } s \leq s_{max} \\ 0 & \text{otherwise} \end{cases} \tag{7a}$$

where $c_{jk}(s) = f_j^*(s) f_k(s)/I_{\text{A}}(s)$ . For the diffraction-difference signal $\Delta I_M(\boldsymbol{s}, t)$ in time-resolved experiments, $\mathcal{M}(\boldsymbol{s})$ becomes

$$\mathcal{M}(\boldsymbol{s}) = \begin{cases} \sum_{j=1}^{N} \sum_{k \neq j}^{N} c_{jk}(s) \iint \left[ e^{-i\boldsymbol{s} \cdot \tilde{\boldsymbol{r}}_{jk}} \tilde{g}_{jk}(\alpha) - e^{-i\boldsymbol{s} \cdot \boldsymbol{r}_{jk}} g_{jk}(\alpha) \right] \sin\alpha \, d\alpha d\beta & \text{for } s \leq s_{max} \\ 0 & \text{otherwise} \end{cases} \tag{7b}$$

where $\tilde{r}_{jk}$, $\tilde{g}_{jk}(\alpha)$ denote the new interatomic distances and angular distributions after excitation, and $r_{\mathrm{jk}}$, $g_{jk}(\alpha)$ denote the interatomic distances and angular distributions before laser excitation. The transform pair between the momentum-transfer and real-space domains is given by

$$\mathcal{P}(r_x, r_y) = \mathcal{A}^{-1}\mathcal{F}_{2\mathcal{D}}^{-1}\left[\mathcal{M}(\boldsymbol{s})e^{-ds^2}\right] \tag{8a}$$

$$\mathcal{M}(\boldsymbol{s})e^{-ds^2} = \mathcal{F}_{2\mathcal{D}}\,\mathcal{A}\left[\mathcal{P}(r_x, r_y)\right] \tag{8b}$$

Where $e^{-ds^2}$ is a damping function used to suppress artifacts caused by truncation at $s_{max}$, and $\mathcal{F}_{2\mathcal{D}}$ and $\mathcal{A}$ denote the forward Fourier and Abel transforms, respectively. The modified pair distribution function $\mathcal{P}(r_x, r_y)$ can also be expressed in polar coordinates $\mathcal{P}(r, \alpha)$, where $r = \sqrt{r_x^2 + r_y^2}$ and $\alpha = \tan^{-1}(r_x/r_y)$.

When $\mathcal{M}(\boldsymbol{s})$ is given by Eq. (7a), the corresponding real-space distribution $\mathcal{P}(r, \alpha)$ can be written as $\sum_{j=1}^{N}\sum_{k=1,j\neq k}^{N} g_{jk}(\alpha)\, h(r - r_{jk}) \otimes \frac{F_j(r)\star F_k(r)}{{r_{jk}}^2}$. Here $h(r - r_{jk})$ is the measured PDF, given by the convolution of $\delta(r - r_{jk})$ with the Fourier transform of the damping function $e^{-ds^2}$ and the Fourier transform of the truncation function associated with the finite detector range.

We consider the case where the diffraction signal is unavailable in the range from 0 Å$^{-1}$ to $s_{min}$ and is approximated by an initial guess function $\mathcal{G}(\boldsymbol{s})$.The corresponding rescaled scattering intensity $\mathcal{M}_e(\boldsymbol{s})$ is defined as

$$\mathcal{M}_e(\boldsymbol{s}) = \begin{cases} \mathcal{G}(\boldsymbol{s}) & \text{for } s < s_{min} \\ \mathcal{M}(\boldsymbol{s}) & \text{for } s_{min} \leq \mathrm{s} \leq s_{max} \\ 0 & \text{otherwise} \end{cases} \tag{9}$$

Analogous to Eq. (8), the transform pair for the measured signal, which includes incorrect information in the region $s < s_{min}$, is defined as

$$\mathcal{P}_e(r, \alpha) = \mathcal{A}^{-1}\mathcal{F}_{2\mathcal{D}}^{-1}\left[\mathcal{M}_e(\boldsymbol{s})e^{-ds^2}\right] \tag{10a}$$

$$\mathcal{M}_e(\boldsymbol{s})e^{-ds^2} = \mathcal{F}_{2\mathcal{D}}\,\mathcal{A}[\mathcal{P}_e(r, \alpha)] \tag{10b}$$

The reconstructed real-space signal $\mathcal{P}_e(r, \alpha)$ can be decomposed as

$$\mathcal{P}_e(r, \alpha) = \mathcal{P}(r, \alpha) + \mathcal{E}(r, \alpha), \tag{11}$$

where $\mathcal{P}(r, \alpha)$ is the true signal that we are seeking, and $\mathcal{E}(r, \alpha)$ is an artifact arising from the difference between $\mathcal{M}_e(\boldsymbol{s})$ and $\mathcal{M}(\boldsymbol{s})$ in the range from 0 Å$^{-1}$ to $s_{min}$. The $\mathcal{E}(r, \alpha)$ can be expressed as

$$\mathcal{E}(r, \alpha) = \mathcal{A}^{-1}\mathcal{F}_{2\mathcal{D}}^{-1}\left[(\mathcal{M}_e - \mathcal{M})e^{-ds^2}\right]\ . \tag{12}$$

Also, we have the following relation

$$[\mathcal{M}_e(\boldsymbol{s}) - \mathcal{M}(\boldsymbol{s})]e^{-ds^2} = \mathcal{F}_{2\mathcal{D}}\,\mathcal{A}[\mathcal{E}(r, \alpha)]. \tag{13}$$

The goal of this work is to iteratively suppress the artifact term $\mathcal{E}(r, \alpha)$ and thereby recover the true scattering signal $\mathcal{M}(\boldsymbol{s})$, using *a priori* knowledge of the minimum and maximum interatomic distances. Successful retrieval relies on two key conditions: (a) the spatial distribution of $\mathcal{E}(r, \alpha)$, associated with missing low-$s$ data ($0 \leq s < s_{\mathrm{min}}$), is broader than that of $\mathcal{P}(r, \alpha)$; (b) approximate bounds on the minimum and maximum interatomic distances of the molecule (and possible reaction products) are available, either from prior knowledge or estimated directly from the data. In general, it is difficult to establish universal requirements on the accessible $s$-range for successful reconstruction, since the constraints are applied in real space and depend on the specific molecular structure and reaction dynamics.

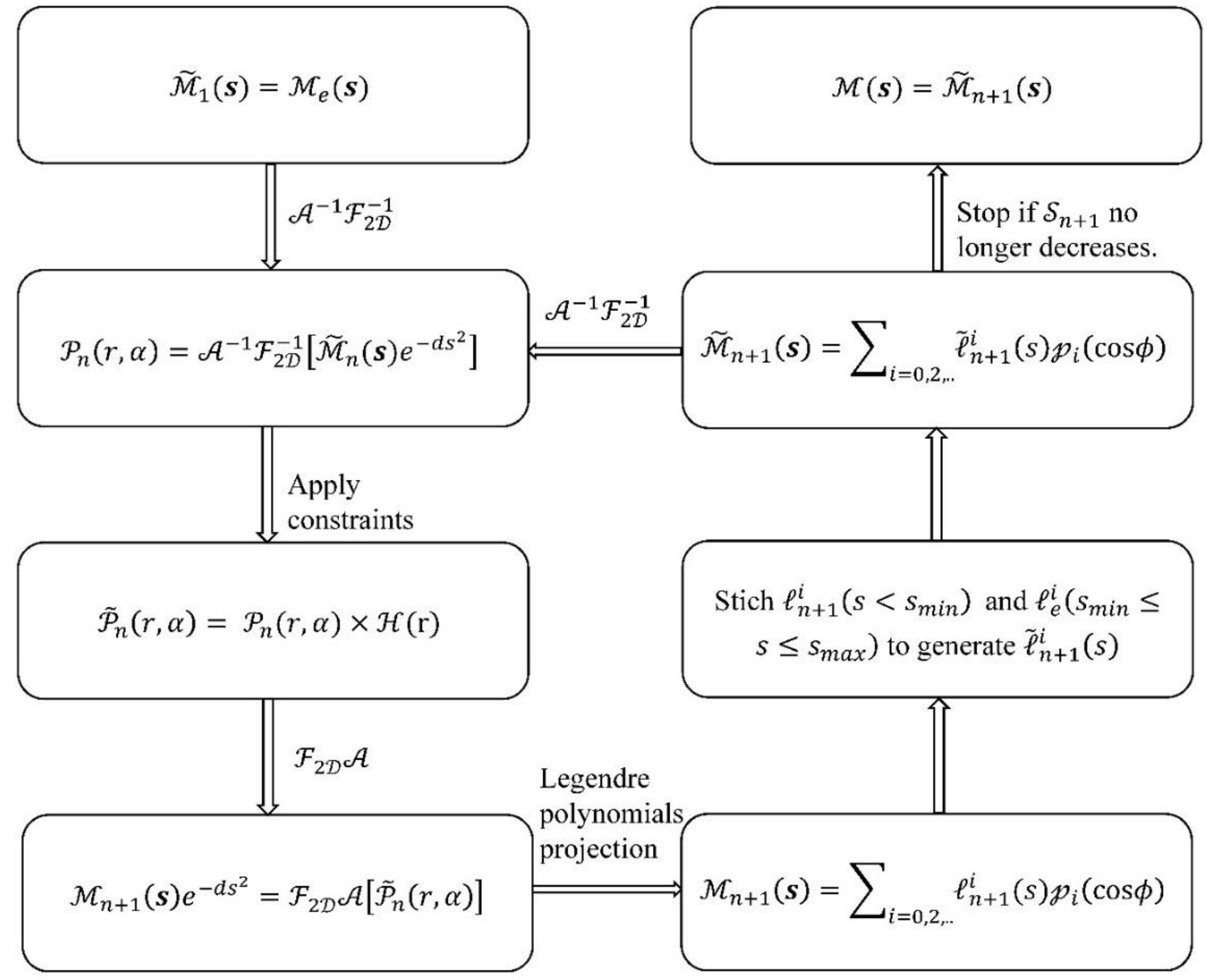

Figure 1. Block diagram of the iterative algorithm used to restore missing low-$s$ signals in 2-D diffraction patterns. The iteration index is denoted by $n$.

Figure 1 illustrates the iterative reconstruction algorithm, with iteration index $n$. The procedure is as follows. (1) At the start, for $n$=1, we set the initial estimate as $\widetilde{\mathcal{M}}_1(\boldsymbol{s}) = \mathcal{M}_e(\boldsymbol{s})$. (2) We apply the Fourier inversion followed by an Abel inversion to obtain $\mathcal{P}_n(r,\alpha) = \mathcal{A}^{-1}\mathcal{F}_{2D}^{-1}\left[\widetilde{\mathcal{M}}_n(\boldsymbol{s})e^{-ds^2}\right]$. (3) We apply a support constraint based on prior knowledge of the molecular structure: $\tilde{\mathcal{P}}_n(r,\alpha) = \mathcal{P}_n(r,\alpha) \times \mathcal{H}(r)$, where $\mathcal{H}(r)$ is a radial band-pass filter that selects the signal within the range $r_1 \leq r \leq r_2$ defined as

$$\mathcal{H}(r) = e^{-\left(\frac{r-r_c}{w}\right)^{2\mathcal{N}}}, \tag{14}$$

where $r_c = (r_1 + r_2)/2$ , $w = (r_2 - r_1)/2$, and $\mathcal{N}$ is a positive integer controlling the sharpness. Since $\mathcal{P}(r,\alpha)$ is confined within this constraint function while the artifact $\mathcal{E}(r,\alpha)$ extends beyond it, this step suppresses the artifact at each iteration. (4) We apply the forward Abel transform followed by the Fourier transform: $\mathcal{M}_{n+1}(\boldsymbol{s})e^{-ds^2} = \mathcal{F}_{2D}\mathcal{A}\left[\tilde{\mathcal{P}}_n(r,\alpha)\right]$. (5) We decompose $\mathcal{M}_{n+1}(\boldsymbol{s})$ into Legendre polynomials [25]:

$$\mathcal{M}_{n+1}(\boldsymbol{s}) = \textstyle\sum_{i=0,2,..} \ell^i_{n+1}(s)\wp_i(\cos\phi), \tag{15}$$

where $\phi$ is the azimuthal angle on the detector plane, and $\wp_i(\cos\phi)$ is the ith order Legendre polynomial. For a linearly polarized laser, only even orders are considered due to the cylindrical symmetry. (6) We combine measured and reconstructed components. The decomposed components of $\mathcal{M}_e(\boldsymbol{s})$ are $\ell_e^i(s)$. The $\ell_{n+1}^i(s < s_{min})$ and $\ell_e^i(s_{min} \le s \le s_{max})$ are stitched together to produce $\tilde{\ell}_{n+1}^i(s)$, in which the signal from 0 to $s_{min}$ is closer to the true signal compared to the previous iteration. To ensure continuity at $s_{\min}$, rescale the reconstructed portion $\ell_{n+1}^i(s < s_{min})$ using a local normalization factor, given by $\int_{s_{min}}^{s_{min}+\mathscr{s}} \ell_e^i(s)ds / \int_{s_{min}-\mathscr{s}}^{s_{min}} \ell_{n+1}^i(s)ds$, where $\mathscr{s}$ is a small value. (7) We generate the updated diffraction pattern $\widetilde{\mathcal{M}}_{n+1}(\boldsymbol{s})$ using components $\tilde{\ell}_{n+1}^i(s)$ and Eq. (15). (8) Replace $\widetilde{\mathcal{M}}_n(\boldsymbol{s})$ with $\widetilde{\mathcal{M}}_{n+1}(\boldsymbol{s})$ to generate $\mathcal{P}_{n+1}(r,\alpha)$ and $\tilde{\mathcal{P}}_{n+1}(r,\alpha)$, and repeat steps (2) – (7) until convergence.

The retrieval error is defined as the sum of the squared differences between the reconstructed $\ell_n^i(s)$ and measured $\ell_e^i(s)$ from $s_{min}$ to $s_{max}$:

$$\mathcal{S}_n = \sum_{i=0,2,..} \int_{s_{min}}^{s_{max}} \left[\ell_n^i(s) - \ell_e^i(s)\right]^2 ds. \tag{16}$$

According to Eq. (13), $\widetilde{\mathcal{M}}_n(\boldsymbol{s})$ converges to the true signal $\mathcal{M}(\boldsymbol{s})$ as the artifact term $\mathcal{E}(r,\alpha)$ approaches zero. The iteration is terminated when $\mathcal{S}_n$ decreases to a small number and reaches a plateau, indicating convergence.

## III. TEST WITH A SIMULATED PATTERN

In this section, we test the retrieval algorithm using simulated diffraction patterns of $CF_3I$. The electron kinetic energy is 90 keV, and atomic scattering amplitudes are taken from tabulated data [42]. To reflect typical experimental conditions, we consider molecular alignment induced by a linearly polarized laser pulse, which produces cylindrical symmetry about the laser polarization axis. For simplicity, the angular distribution of the C–I bond is modeled as $g_{\mathrm{CI}}(\alpha) = \frac{3}{2}\cos^2\alpha$, while the angular distributions for other atom pairs, $g_{jk}(\alpha)$, can be obtained using the method described in Ref. [31].

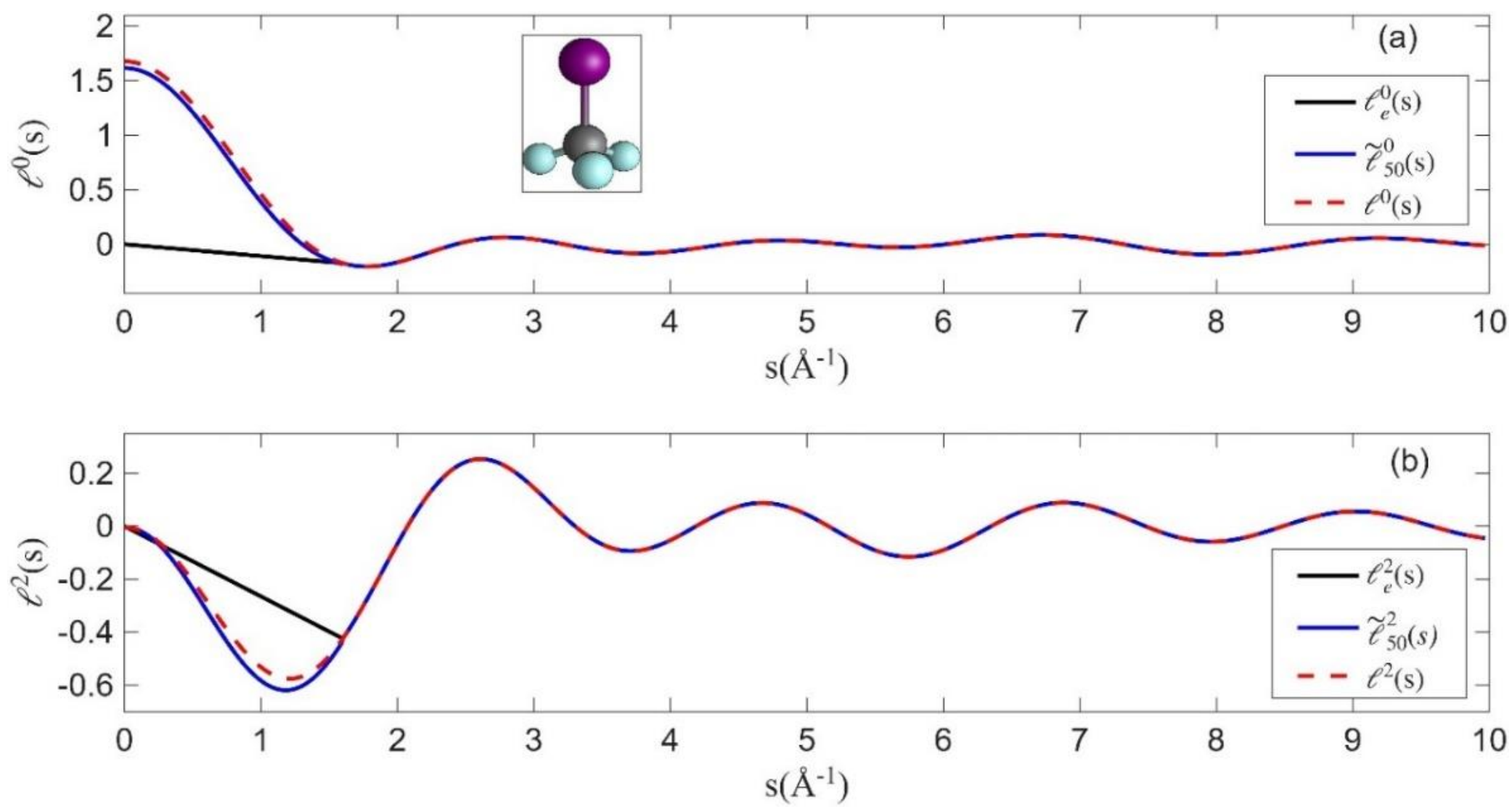

Figure 2. Input and reconstructed Legendre components $\ell_n^i(s)$. (a) Zeroth-order component: the initial approximation $\ell_e^0(s)$ (solid black) is obtained by linear interpolation in the missing low-$s$ region; the reconstructed signal $\tilde{\ell}_{50}^0(s)$ after 50 iterations is shown in solid blue; the true signal $\ell^0(s)$ is shown as a dashed red line. The inset shows the molecular structure of $CF_3I$, with carbon (gray), iodine (purple), and fluorine (green). (b) Second-order component $\ell_e^2(s)$ (solid black), reconstructed $\tilde{\ell}_{50}^2(s)$(solid blue), and true signal $\ell^2(s)$ (dashed red).

The molecular structure of $CF_3I$ molecule [43] is shown in Fig. 2(a), where iodine, carbon, and fluorine atoms are represented in purple, gray, and green, respectively. The minimum and maximum interatomic distances are $r_{\mathrm{CF}} = 1.33$Å and $r_{\mathrm{FI}} = 2.89$Å. Based on this, the real-space constraint function $\mathcal{H}(\mathrm{r})$ is defined with $r_1 = 1.15$Å, $r_2 = 3.20$Å, and $\mathcal{N} = 15$. The damping parameter is set to $d = 0.015$ Å$^2$. The diffraction signal $I_M(\boldsymbol{s})$ and rescaled signal $\mathcal{M}(\boldsymbol{s})$ are calculated using Eqs. (4) and (7a). We restrict the accessible momentum-transfer range to 1.6 Å$^{-1}$ $\leq s \leq$ 10 Å$^{-1}$, which is consistent with typical GUED experiments [3, 31, 33, 44]. The diffraction pattern with the missing low-*s* region is decomposed into Legendre components $\ell_e^i(s)$ with $i = 0{,}2$, while higher-order terms are neglected due to their negligible amplitudes. Higher orders can be included in the Legendre projection if the amplitudes are not small. The missing region 0 Å$^{-1}$ $\leq s \leq$ 1.6 Å$^{-1}$ is initially filled using linear interpolation, formulated as $s\ell_e^i(s_{min})/s_{min}$, which corresponds to $\mathcal{G}(\boldsymbol{s})$. The $\ell_e^0(s)$ and $\ell_e^2(s)$ are shown as solid black lines in Figs. 2(a) and 2(b). The iterative algorithm simultaneously reconstructs the Legendre components $\ell_n^i(s)$ and the corresponding diffraction patterns (Fig. 3). After 50 iterations, the retrieved components $\ell_{50}^i(s)$ (solid blue lines) closely match the true signals $\ell^i(s)$ (dashed red lines), as shown in Figs. 2(a) and 2(b). In particular, the reconstructed low-$s$ region (0 Å$^{-1}$ $\leq s \leq$ 1.6 Å$^{-1}$) shows excellent agreement with the true signal, demonstrating successful recovery of the missing data.

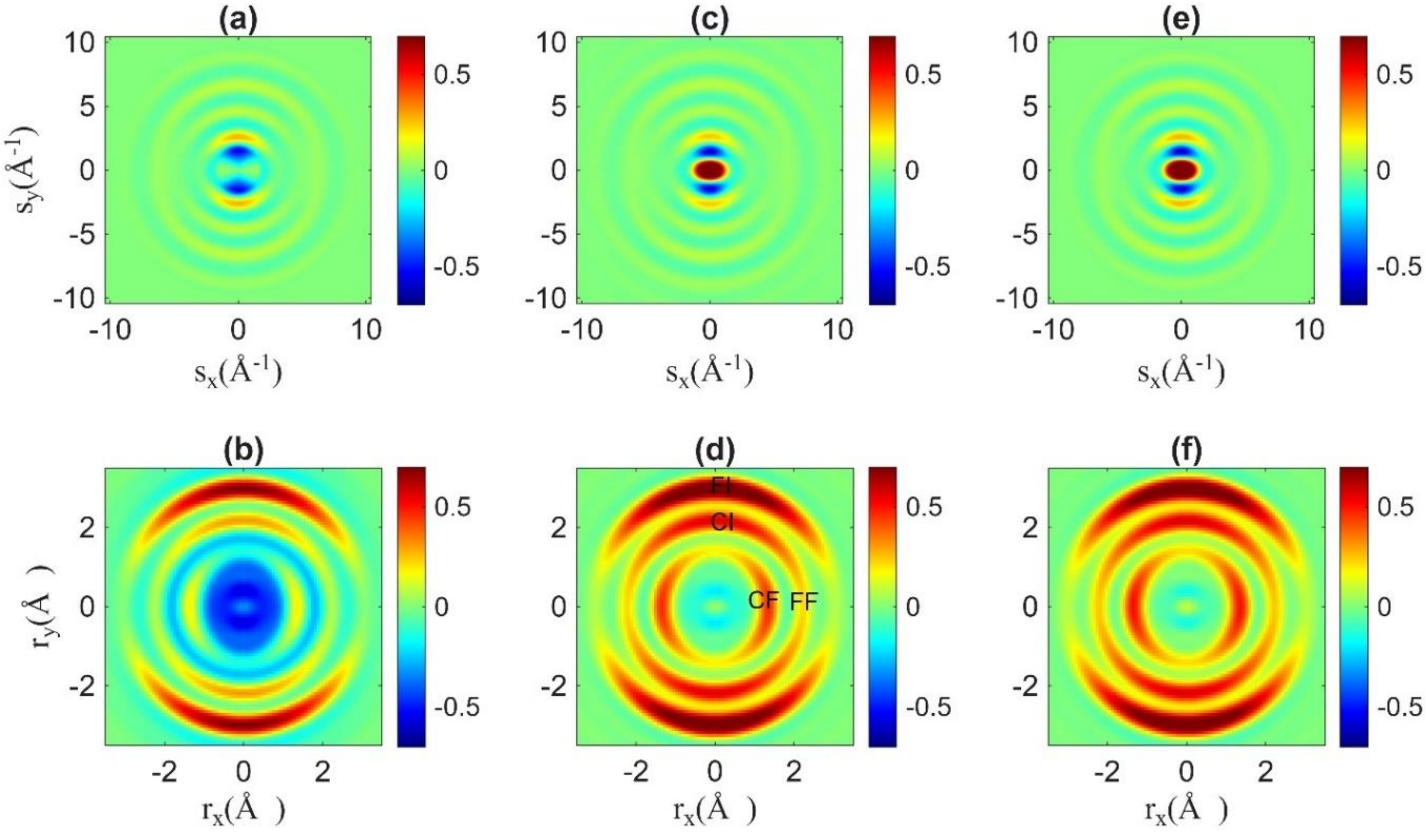

Figure 3. Input and reconstructed diffraction patterns of $CF_3I$. (a) Initial diffraction pattern $\widetilde{\mathcal{M}}_1(\boldsymbol{s})$, which corresponds to Legendre components $\ell_e^i(s)$ with $i = 0{,}2$ shown in Fig. 2. (b) Corresponding real-space distribution $\mathcal{P}_1(r, \alpha)$, showing strong artifacts due to missing low-$s$ data. (c) Reconstructed diffraction pattern $\widetilde{\mathcal{M}}_{50}(\boldsymbol{s})$ after 50 iterations. (d) Real-space distribution $\mathcal{P}_{50}(r, \alpha)$, where the artifacts are largely removed. The rings correspond to interatomic distances $r_{\mathrm{CF}} = 1.33$Å, $r_{\mathrm{CI}} = 2.14$Å, $r_{\mathrm{FF}} = 2.15$Å, and $r_{\mathrm{FI}} = 2.89$Å. (e) True diffraction pattern $\mathcal{M}(\boldsymbol{s})$. (f) MPDF of the true diffraction pattern $\mathcal{P}(r, \alpha)$.

The input and reconstructed diffraction patterns are shown in Fig. 3. In the first iteration, the diffraction pattern $\widetilde{\mathcal{M}}_1(\boldsymbol{s})$ is constructed from the initial Legendre components $\ell_e^0(s)$ and $\ell_e^2(s)$, shown in Fig. 2, formulated as $\widetilde{\mathcal{M}}_1(\boldsymbol{s}) = \sum_{i=0,2} \ell_e^i(s) \wp_i(\cos\phi)$, and is shown in Fig. 3(a). The corresponding real-space distribution $\mathcal{P}_1(r,\alpha)$ , obtained via inverse transforms, is shown in Fig. 3(b). Due to the missing low-$s$ data, significant artifacts are present in $\mathcal{P}_1(r,\alpha)$. After 50 iterations, the reconstructed diffraction pattern $\widetilde{\mathcal{M}}_{50}(\boldsymbol{s})$ and real-space distribution $\mathcal{P}_{50}(r,\alpha)$, shown in Figs. 3(c) and 3(d), are in good agreement with the true signals $\mathcal{M}(\boldsymbol{s})$ and $\mathcal{P}(r,\alpha)$, shown in Figs. 3(e) and 3(f). The pair distribution functions in Fig. 3(d) show distinct rings corresponding to interatomic distances $r_{\mathrm{CF}} = 1.33$Å, $r_{\mathrm{CI}} = 2.14$Å, $r_{\mathrm{FF}} = 2.15$Å, and $r_{\mathrm{FI}} = 2.89$Å. The angular intensity variation along each ring reflects the corresponding atom-pair angular distribution.

The retrieval error for the simulated diffraction pattern is quantified using the sum of the squared residuals in the missing region ($s < s_{min}$) between the reconstructed components $\ell_n^i(s)$ and the true signal $\ell^i(s)$:

$$\mathcal{R}_n = \sum_{i=0,2,..} \int_0^{s_{min}} \left[\ell_n^i(s) - \ell^i(s)\right]^2 ds. \tag{17}$$

Figure 4 shows both error metrics: $\mathcal{S}_n$(left axis) and $\mathcal{R}_n$(right axis). While $\mathcal{R}_n$ is only accessible for simulated data, $\mathcal{S}_n$ can be evaluated for experimental data where the true signal is unknown. Both error metrics $\mathcal{S}_n$ and $\mathcal{R}_n$ exhibit similar convergence behavior, both decreasing and reaching a plateau after approximately 15 iterations.

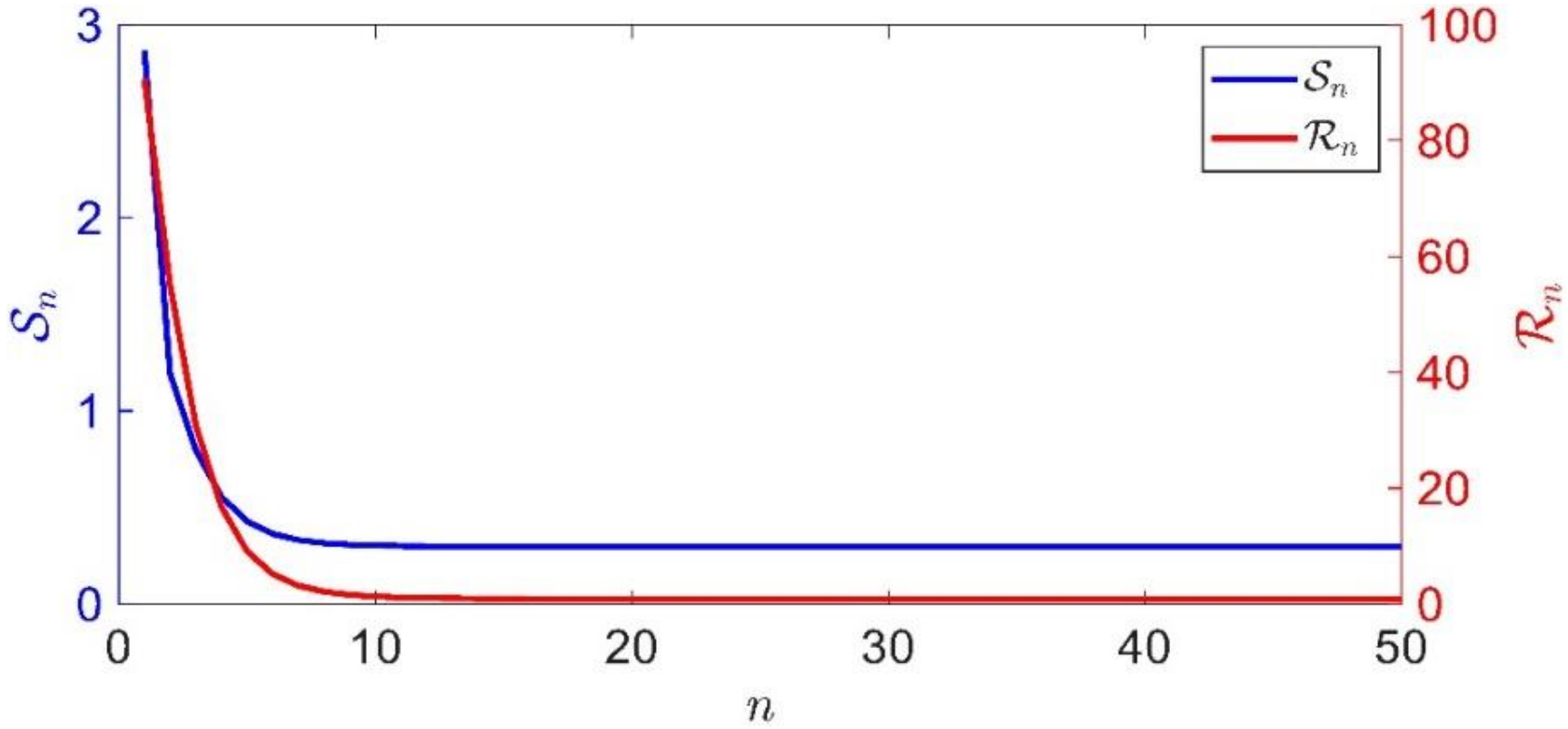

Figure 4. The error metrics $\mathcal{S}_n$ and $\mathcal{R}_n$ as a function of iteration number $n$. The iteration number is denoted as $n$.

## IV. APPLICATION TO EXPERIMENTAL PATTERN

We apply the iterative reconstruction algorithm to experimental diffraction patterns of $CF_3I$ undergoing impulsive alignment induced by a femtosecond laser pulse. The measurements were performed using a tabletop keV-UED instrument. The detail of the experiment is available in [31]. An 800 nm femtosecond infrared laser pulse generates a rotational wave packet, and time-resolved diffraction patterns are recorded to capture the alignment dynamics. Shortly after excitation, a prompt alignment is observed, producing anisotropic diffraction patterns, in contrast to the circularly symmetric patterns of randomly oriented molecules. To evaluate the performance of the algorithm in the presence of experimental noise, we focus on reconstructing the missing low-$s$ signal at the prompt alignment peak.

The diffraction-difference signal $\Delta I_M(\boldsymbol{s})$ is obtained by subtracting the pre-excitation pattern from that at the alignment peak. The molecular scattering intensity is then reconstructed as $I_M(\boldsymbol{s}) = \Delta I_M(\boldsymbol{s}) + a \cdot$ $I_M^{randon}(s)$, where $I_M^{randon}(s)$ is the theoretical scattering signal for randomly oriented molecules. The scale factor $a = 0.28$ is determined by fitting the experimental data to theoretical calculations (see Ref. [31] for details). The $I_M(\boldsymbol{s})$ is used to define the measured signal $\mathcal{M}_e(\boldsymbol{s})$ over the accessible range $1.6\ \text{Å}^{-1} \leq s \leq 10\ \text{Å}^{-1}$, following Eq. (9). The diffraction pattern is decomposed into Legendre components $\ell_e^i(s)$ with $i = 0, 2$, and the missing region $0\ \text{Å}^{-1} \leq s < 1.6\ \text{Å}^{-1}$ is initialized using a linear interpolation.

The input and reconstructed Legendre components $\ell_n^i(s)$ and the corresponding 2-D diffraction patterns are shown in Figs. 5 and 6. The parameters of the constraint function $\mathcal{H}(\mathrm{r})$ and the damping constant are the same as those used in Sec. III. The initial components $\ell_e^0(s)$ and $\ell_e^2(s)$ are shown as solid black lines in Figs. 5(a) and 5(b), respectively. The reconstructed components after 50 iterations, $\tilde{\ell}_{50}^i(s)$, are shown as solid blue lines. For comparison, theoretical molecular scattering signals $I_M(\boldsymbol{s})$ are calculated using atom-pair angular distributions $g_{jk}(\alpha)$ obtained from numerical solutions of the time-dependent Schrödinger equation (TDSE), based on the experimental laser parameters and rotational temperature [45]. The Legendre components $\ell^i(s)$ from theoretical signal are shown as dashed red lines in Figs. 5(a) and 5(b). These theoretical curves are included only for reference and are not used in the reconstruction of the experimental data. More details of the calculations are provided in Ref. [31].

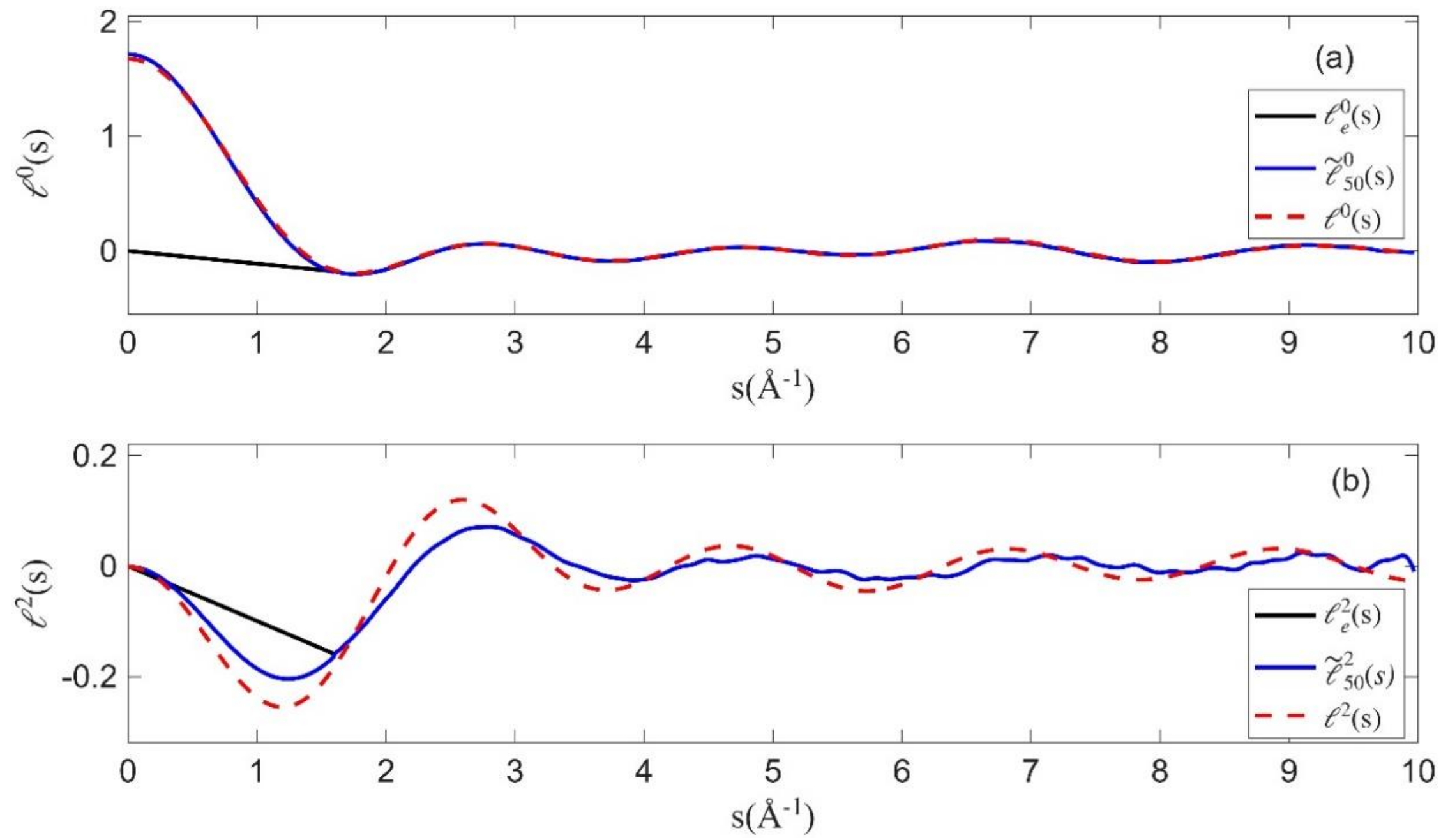

Figure 5. Input and reconstructed Legendre components $\ell_n^i(s)$ for experimental data. (a) Zeroth-order component: the initial estimate $\ell_e^0(s)$ (solid black) is obtained by linear interpolation in the missing low-$s$ region; the reconstructed result $\tilde{\ell}_{50}^0(s)$ after 50 iterations is shown in solid blue; the theoretical signal $\ell^0(s)$ is shown as a dashed red line. (b) Second-order component: $\ell_e^2(s)$ (solid black), reconstructed $\tilde{\ell}_{50}^2(s)$ (solid blue), and theoretical $\ell^2(s)$ (dashed red).

The input and reconstructed 2-D diffraction patterns are shown in Fig. 6. The initial pattern $\widetilde{\mathcal{M}}_1(\boldsymbol{s})$ is shown in Fig. 6(a), with the corresponding real-space distribution $\mathcal{P}_1(r, \alpha)$ in Fig. 6(b). Due to the missing low-$s$ data, strong artifacts are observed in $\mathcal{P}_1(r, \alpha)$. After 50 iterations, the reconstructed diffraction pattern $\widetilde{\mathcal{M}}_{50}(\boldsymbol{s})$ and real-space distribution $\mathcal{P}_{50}(r, \alpha)$, shown in Figs. 6(c) and 6(d), show substantial improvement, with the artifacts largely suppressed. These results are in good agreement with the theoretical diffraction pattern $\mathcal{M}(\boldsymbol{s})$ and its real-space representation $\mathcal{P}(r, \alpha)$, shown in Figs. 6(e) and 6(f). The convergence behavior is shown in Fig. 7, where the error metric $\mathcal{S}_n$ decreases rapidly and approaches a minimum after approximately 10 iterations.

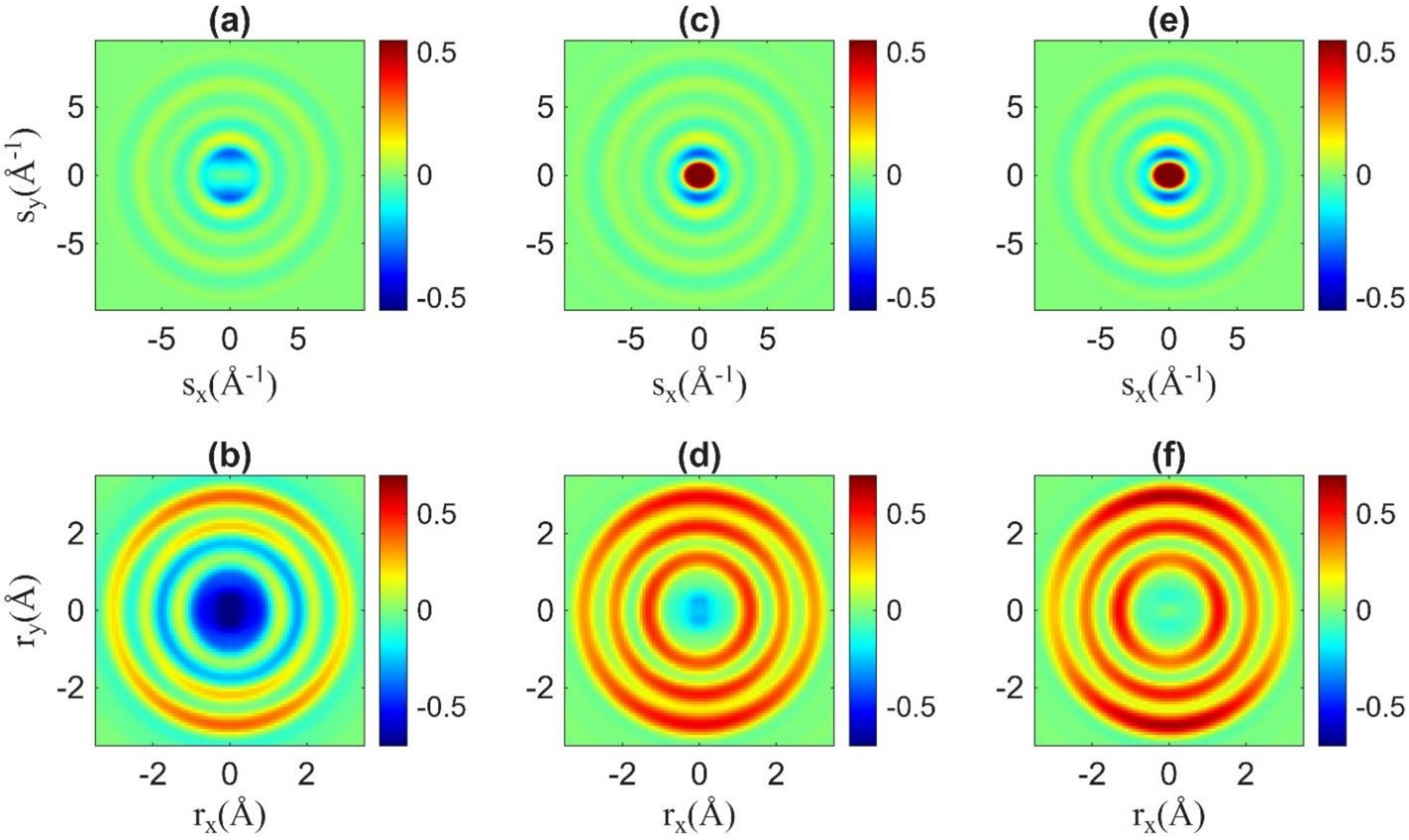

Figure 6. Input and reconstructed experimental diffraction patterns of $CF_3I$ alignment induced by an infrared laser pulse. (a) Initial diffraction pattern $\widetilde{\mathcal{M}}_1(\boldsymbol{s})$. (b) Corresponding real-space distribution $\mathcal{P}_1(r, \alpha)$, showing strong artifacts due to missing low-$s$ data. (c) Reconstructed diffraction pattern $\widetilde{\mathcal{M}}_{50}(\boldsymbol{s})$ after 50 iterations. (d) Real-space distribution $\mathcal{P}_{50}(r, \alpha)$, with artifacts largely suppressed. (e) Theoretical diffraction pattern $\mathcal{M}(\boldsymbol{s})$, calculated from TDSE using the experimental laser parameters and rotational temperature. (f) Real-space distribution $\mathcal{P}(r, \alpha)$.

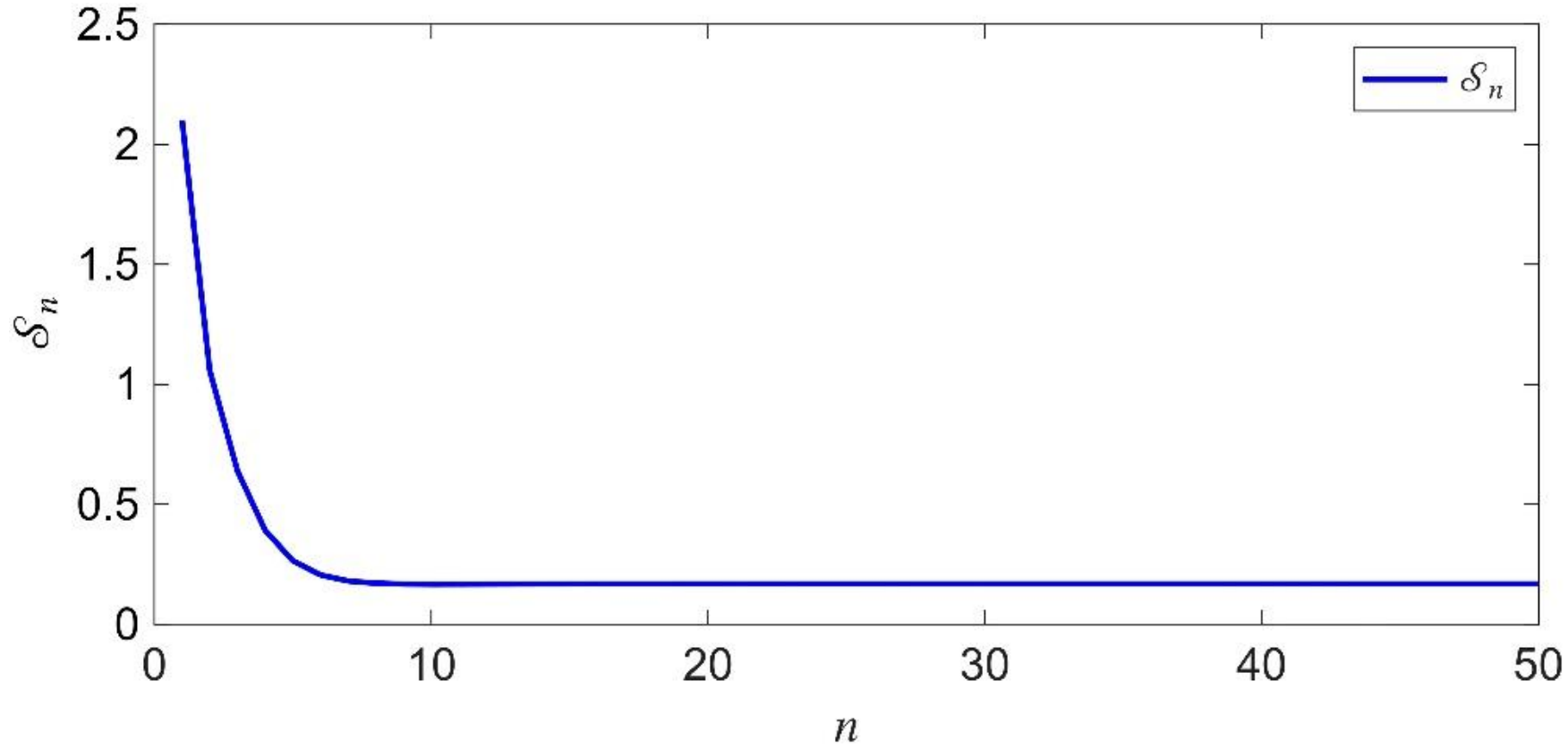

Figure 7. The function $\mathcal{S}_n$ computed in retrieving the impulsive alignment diffraction pattern of $CF_3I$. The iteration number is denoted as $n$.

## V. CONCLUSION

Anisotropic 2-D diffraction signals, commonly observed in GUED and UXRD experiments due to the use of linearly polarized laser pump, contain richer structural information than conventional isotropic scattering. However, missing data at low momentum transfer limits accurate real-space reconstruction, including the determination of interatomic distances and atom-pair angular distributions. In this work, we have developed an iterative algorithm to restore missing low-$s$ signals in 2-D diffraction patterns. The method transforms between the momentum-transfer and real-space domains using coupled Fourier and Abel transforms, while enforcing real-space constraints to suppress artifacts and restore the missing data from experimentally measured diffraction pattern. We demonstrate successful reconstruction of the missing signal in both simulated and experimental GUED patterns of laser aligned $CF_3I$ molecules over typical momentum-transfer ranges. We have also tested the algorithm for more limited $s$-ranges (e.g., $1.6\text{Å}^{-1} \leq s \leq 5.0\text{Å}^{-1}$), typical of UXRD measurements [10, 46, 47], and diffraction pattern in the presence of significant noise (see Appendices A and B).The method also performs well for more limited ranges and remains robust in the presence of significant noise. The algorithm is simple to implement and requires only a minimal *priori* knowledge of the molecular structure, namely approximate bounds on interatomic distances. Moreover, it provides a general framework applicable to both anisotropic and isotropic diffraction patterns, and the latter corresponds to previously reported 1-D signal restoration methods [18]. These results demonstrate that missing low-$s$ information can be reliably recovered from experimentally accessible data, enabling more complete structural characterization in ultrafast diffraction studies.

## ACKNOWLEDGMENTS

This work was supported by the US Department of Energy Office of Science, Basic Energy Sciences under award no. DE-SC0014170.

## DATA AVAILABILITY

The data that support the findings of this article are openly available in [48].

## APPENDIX A: SIGNAL WITH A SMALLER S-RANGE

In Sec. III, we demonstrated that the iterative algorithm can successfully restore the missing low-$s$ region ($0\text{Å}^{-1} \leq s \leq 1.6\text{Å}^{-1}$) using simulated data over the range $1.6\text{Å}^{-1} \leq s \leq 10\text{Å}^{-1}$, corresponding to typical GUED conditions. Here, we further test the performance of the algorithm with a more limited momentum-transfer range. Using the same simulated diffraction pattern of aligned $CF_3I$ described in Sec. III, we restrict the available data to $1.6\text{Å}^{-1} \leq s \leq 5.0\text{Å}^{-1}$ and attempt to reconstruct the missing low-$s$ region. The parameters of the constraint function $\mathcal{H}(\mathrm{r})$ and the damping constant are kept identical to those used in Sec. III. The input signals $\ell_e^0(s)$ and $\ell_e^2(s)$ are shown as solid black lines in Fig. 8(a) and 8(b), respectively. The reconstructed components $\tilde{\ell}_{50}^i(s)$ after 50 iterations (solid blue) show good agreement with the true signals $\ell^i(s)$ (dashed red), with significantly reduced discrepancies in the low-$s$ region. Despite the reduced momentum transfer range, the algorithm accurately recovers the missing low-$s$ signal. The corresponding 2-D diffraction patterns are not shown, as they can be uniquely reconstructed from the Legendre components $\ell_n^i(s)$.

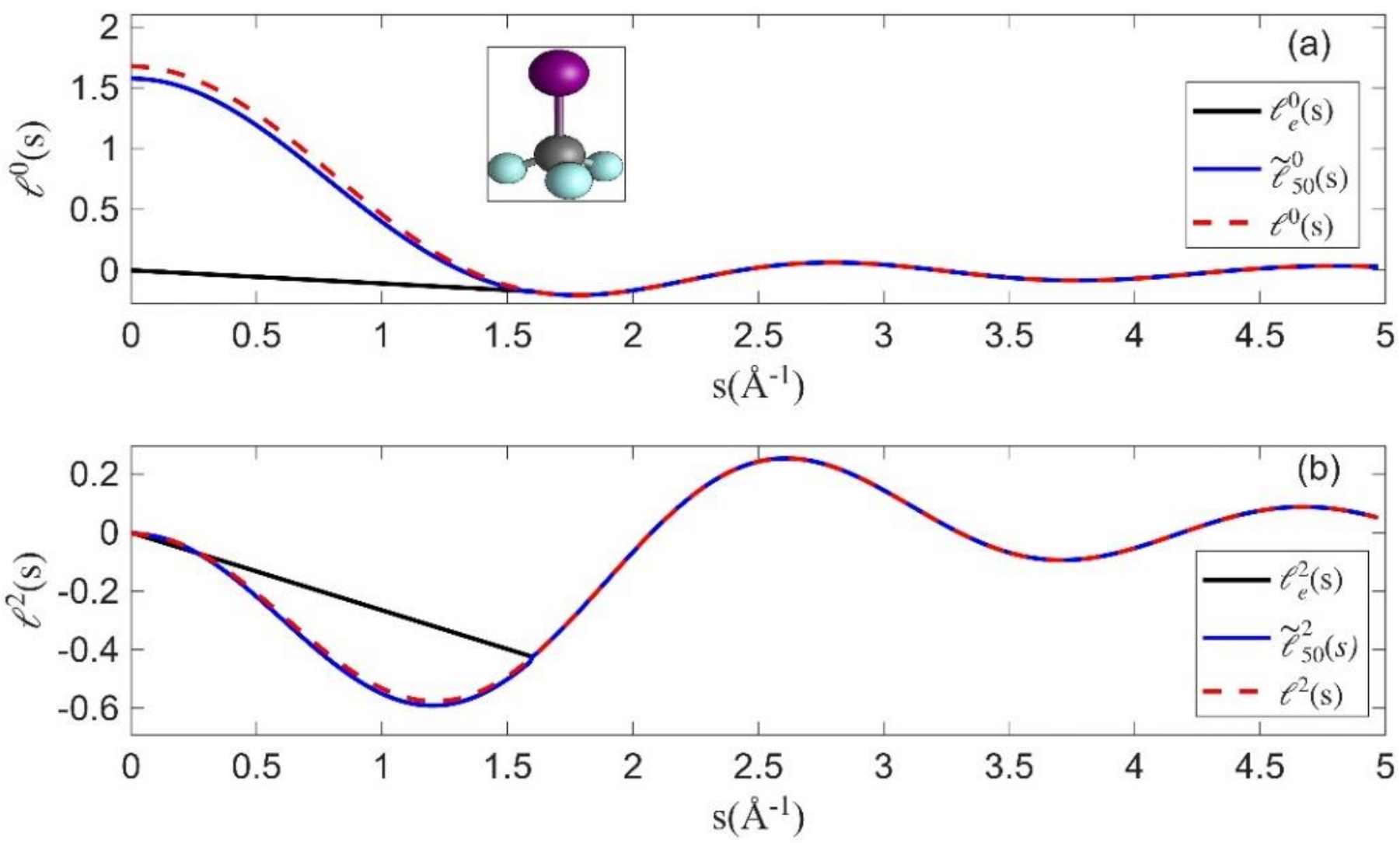

Figure 8. Input and reconstructed Legendre components $\ell_n^i(s)$ for the limited $s$-range case. (a) Zeroth-order component: the initial estimate $\ell_e^0(s)$ (solid black); the reconstructed result $\tilde{\ell}_{50}^0(s)$ after 50 iterations (solid blue); and the true signal $\ell^0(s)$ (dashed red). (b) Second-order component: $\ell_e^2(s)$ (solid black), reconstructed $\tilde{\ell}_{50}^2(s)$ (solid blue), and true signal $\ell^2(s)$ (dashed red).

## APPENDIX B: DATA WITH NOISE

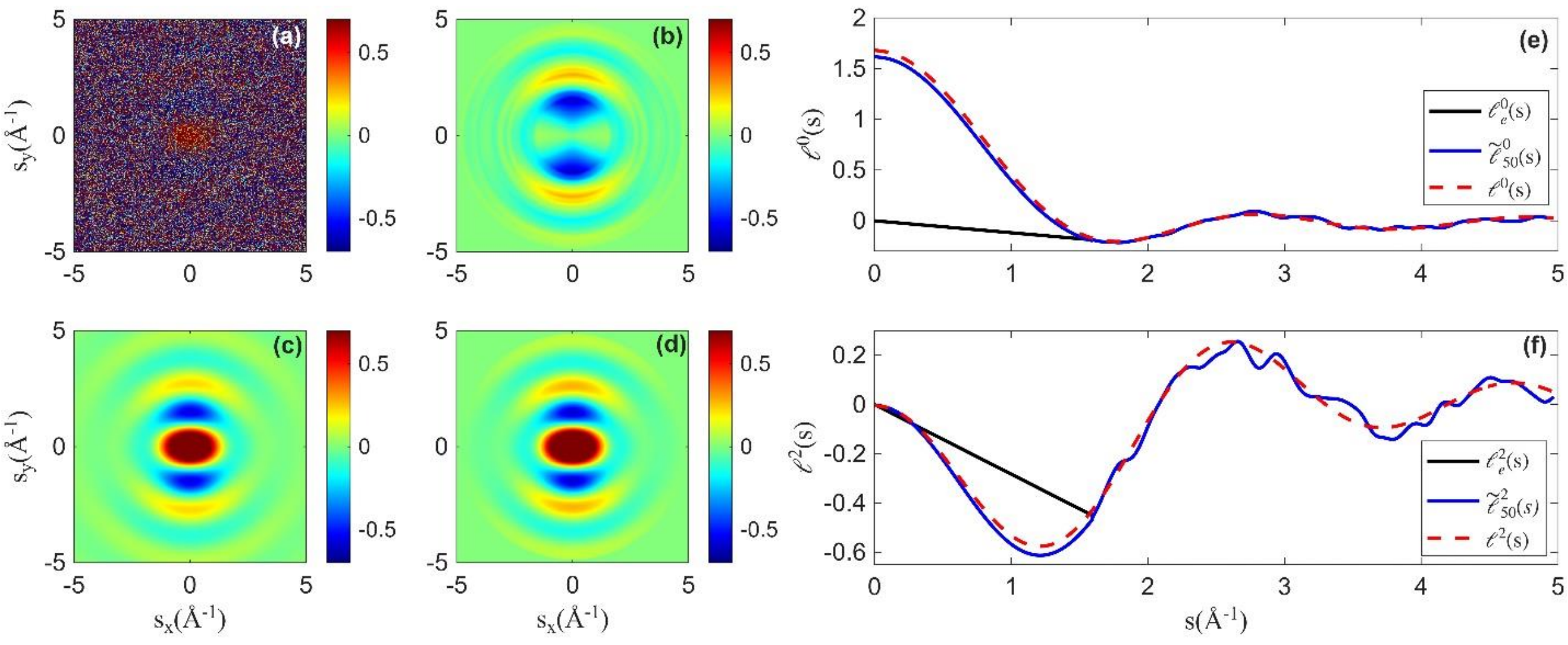

Figure 9. Reconstruction performance in the presence of noise. (a) Simulated diffraction pattern with added random noise, $\mathcal{M}(\boldsymbol{s}) + n(\boldsymbol{s})$. (b) Input pattern $\mathcal{M}_e(\boldsymbol{s})$, reconstructed from the Legendre components $\ell_e^0(s)$ and $\ell_e^2(s)$. (c) Reconstructed pattern $\widetilde{\mathcal{M}}_{50}(\boldsymbol{s})$after 50 iterations. (d) Theoretical diffraction pattern $\mathcal{M}(\boldsymbol{s})$. (e) Zeroth-order Legendre component: initial estimate $\ell_e^0(s)$ (solid black), reconstructed $\tilde{\ell}_{50}^0(s)$(solid blue), and true signal $\ell^0(s)$ (dashed red). (f) Second-order component: $\ell_e^2(s)$ (solid black), reconstructed $\tilde{\ell}_{50}^2(s)$(solid blue), and true signal $\ell^2(s)$(dashed red).

In this section, we evaluate the performance of the iterative algorithm in the presence of significant noise. A random noise pattern $n(\boldsymbol{s})$ is added to $\mathcal{M}(\boldsymbol{s})$ to generate a noisy input signal. The noise is simulated using the MATLAB function randi, which generates a 2-D array of uniformly distributed random integers in the range $[-100,100]$. The noise amplitude is scaled by max $(\mathcal{M})/100$, where max $(\mathcal{M})$denotes the maximum value of $\mathcal{M}(\boldsymbol{s})$. The resulting noisy diffraction pattern $\mathcal{M}(\boldsymbol{s}) + n(\boldsymbol{s})$ is shown in Fig. 9(a). The available momentum-transfer range used for retrieval is $1.6\text{Å}^{-1} \leq s \leq 5.0\text{Å}^{-1}$. The input pattern $\mathcal{M}_e(\boldsymbol{s})$, reconstructed from the Legendre components $l_e^0(s)$and $l_e^2(s)$, is shown in Fig. 9(b), with the components plotted as solid black lines in Figs. 9(e) and 9(f). A Gaussian-weighted moving average with an 11-point window is applied to reduce noise in Legendre components while preserving their key features. The resulting $\mathcal{M}_e(\boldsymbol{s})$ is used as the input for the first iteration. Figures 9(c) and 9(d) show the reconstructed diffraction pattern $\widetilde{\mathcal{M}}_{50}(\boldsymbol{s})$ after 50 iterations and the true pattern $\mathcal{M}(\boldsymbol{s})$, respectively. In Figs. 9(e) and 9(f), the reconstructed components $\tilde{\ell}_{50}^i(s)$(solid blue) show good agreement with the true signals $\ell^i(s)$(dashed red), with significantly reduced discrepancies in the low-$s$ region. The restored data over $0\text{Å}^{-1} \leq s \leq 1.6\text{Å}^{-1}$ closely matches the true signal. Despite strong noise contamination, the algorithm accurately recovers the missing low-$s$ information, demonstrating its robustness, stability, and reliability under noisy conditions.